Towards structured sharing of raw and derived neuroimaging data across existing resources

Keator D.B.[1,2], Helmer K.[3], Steffener J.[4], Turner J.A.[5], Van Erp T.G.M.[1], Gadde S.[6], Ashish N.[7], Burns G.A.[8], Nichols B.N.[9,10], Ghosh, S.S.[11]

[1] Department of Psychiatry and Human Behavior, University of California, Irvine
[2] Department of Computer Science, University of California, Irvine
[3] Athinoula A. Martinos Center for Biomedical Imaging, Massachusetts General Hospital and Department of Radiology, Harvard Medical School, Boston, MA, USA
[4] Taub Institute for Research on Alzheimer's Disease and the Aging Brain, Department of Neurology, Columbia University
[5] The Mind Research Network, Albuquerque, New Mexico
[6] Brain Imaging and Analysis Center, Duke University
[7] California Institute of Telecommunications and Information Technology (CalIT2), University of California, Irvine.
[8] Information Sciences Institute, University of Southern California
[9] Department of Biomedical Informatics and Medical Education, University of Washington, Seattle
[10] Integrated Brain Imaging Center, Department of Radiology, University of Washington, Seattle
[11] McGovern Institute for Brain Research, Massachusetts Institute of Technology, Cambridge, MA

Abstract

Data sharing efforts increasingly contribute to the acceleration of scientific discovery. Neuroimaging data is accumulating in distributed domain-specific databases and there is currently no integrated access mechanism nor an accepted format for the critically important meta-data that is necessary for making use of the combined, available neuroimaging data. In this manuscript, we present work from the Derived Data Working Group, an open-access group sponsored by the Biomedical Informatics Research Network (BIRN) and the International Neuroimaging Coordinating Facility (INCF) focused on practical tools for distributed access to neuroimaging data.  The working group develops models and tools facilitating the structured interchange of neuroimaging meta-data and is making progress towards a unified set of tools for such data and meta-data exchange. We report on the key components required for integrated access to raw and derived neuroimaging data as well as associated meta-data and provenance across neuroimaging resources. The components include (1) a structured terminology that provides semantic context to data, (2) a formal data model for neuroimaging with robust tracking of data provenance, (3) a web service-based application programming interface (API) that provides a consistent mechanism to access and query the data model, and (4) a provenance library that can be used for the extraction of provenance data by image analysts and imaging software developers. We believe that the framework and set of tools outlined in this manuscript have great potential for solving many of the issues the neuroimaging community faces when sharing raw and derived neuroimaging data across the various existing database systems for the purpose of accelerating scientific discovery.

# 1  Introduction

Acceleration of scientific discovery in neuroimaging and many other research areas increasingly relies on the availability of large and well-documented data sets. In fact, many of the major new discoveries in the genetics of schizophrenia and other psychiatric disorders, multiple sclerosis, diabetes, obesity, and other metabolic traits have been possible only through collaborative data sharing (Ripke et al., 2011; Sawcer et al., 2011; Speliotes et al., 2010). In the area of neuroimaging, such data sets can be obtained by a) funding large consortia to prospectively acquire large data sets (Insel et al., 2004), b) harvesting research-ready data from other sources (Kho et al., 2011; van Erp et al., 2011), and/or c) data (Biswal et al., 2010) or analysis results (Stein et al., 2012), sharing between multiple separately funded initiatives that include in-common measurements.  In-common measurements, in the context of neuroimaging, refer to imaging protocols that are included in many magnetic resonance imaging (MRI) related studies such as resting state functional magnetic resonance imaging (fMRI), structural T1-weighted MRI, and diffusion tensor imaging (Nooner et al., 2012).  Shared and combined use of in-common measurements is the lowest barrier in the otherwise complex and often intractable space of combining neuroimaging data collected under different initiatives; however, acquiring equivalent data sets at sites with hardware from different vendors requires careful protocol design (Jack et al., 2010; Jack et al., 2008; Kruggel et al., 2010).  Despite efforts from consortia such as the Function and Morphometry test beds of the Biomedical Informatics Research Network (BIRN) that have published recommendations for collecting neuroimaging data with the sharing and combining of data from multiple sites in mind, the task of data sharing across scanner platforms remains difficult even though the benefits are both financially and scientifically undeniable (Glover et al., 2012; Poline et al., 2012). Sharing well-documented, often publicly funded, data sets for use by the wider research community can be cost-effective as it allows for 1) increased statistical power through mega-analyses in contrast to meta-analyses, 2) obtaining new larger data sets to answer questions not addressed by the original studies, 3) application of newly developed tools to existing data sets, and 4) replication of research findings via reanalysis of existing data by other research groups.

In the last ten years, large neuroimaging datasets have become publicly available, although, there are significant differences in the requirements for data access.  These data sets are in domain-specific repositories.  Some examples of completely open-access neuroimaging repositories include XNAT Central (central.xnat.org) which includes over 3000 subjects stored in the XNAT database (Marcus et al., 2007), the BIRN data repository (www.birncommunity.org/resources/data) which includes large cohorts of both mouse and human imaging data stored in the BIRN Human Imaging Database (Florescu et al., 1996; Ozyurt et al., 2010)) and elsewhere,  the 1000 Functional Connectomes project (www.nitrc.org/projects/fcon_1000/) which, at the present time, contains over 1000 subjects, and the relatively new OpenFMRI repository (www.openfmri.org) which contains imaging data from over 200 subjects.  The Neuroimaging Informatics Tools and Resources Clearinghouse (NITRC, www.nitrc.org) hosts neuroimaging data, in addition to neuroimaging processing and analysis tools (Buccigrossi et al., 2008).  Examples of neuroimaging repositories that require some form of permission to download data (e.g. prior IRB approval or simply an application to the host site), include the Alzheimer's Disease Neuroimaging Initiative (ADNI; http://adni.loni.ucla.edu/) which contains imaging data from over 800 subjects, and the National Database for Autism Research (NDAR; ndar.nih.gov) which contains data from over 6000 subjects.  It is clear from this short (and by no means exhaustive) list of available neuroimaging repositories that data is accumulating in distributed domain-specific databases, rather than in a small number of central repositories.  In addition, there is no integrated access mechanism, even for open-access resources, nor an accepted format for the critically important meta-data, necessary for making use of combined neuroimaging data.  The Neuroscience Information Framework (NIF;www.neuinfo.org) (Gupta et al., 2008) provides integrated access to many neuroscience-related databases as well as other resources; researchers can identify imaging datasets for download from certain resources that have been mapped for the NIF interface, for example. Developing the meta-data formats and standards needed to understand the imaging datasets, or to capture the details of how the data were collected and processed, is outside the scope of NIF and other database mediators. Integrated access to existing resources, many already identified by NIF, when combined with such meta-data documentation, would provide a full-service shop for queries and download of publicly available data across projects.

A critical barrier in enabling structured sharing of raw and derived neuroimaging data across existing resources is the lack of a standard meta-data model and a set of informatics tools that enables the sharing of meta-data, including provenance, associated with neuroimaging data (Teeters et al., 2008).  Meta-data are descriptive elements associated with data that provide additional clarity regarding acquisition parameters, experimental conditions, analysis procedures, and any other formation about the experiment or analyses that helps one understand and use the data. The  benefits of neuroimage data sharing was introduced more than a decade ago (Van Horn et al., 2001; Van Horn and Gazzaniga, 2002), several successful data sharing projects exist (Biswal et al., 2010; Weiner et al., 2012; Nooner et al., 2012), and many of the technical, legal, and social issues of data sharing have been discussed (Poline et al., 2012; Milham, 2012; Mennes et al., 2012), but there currently is no standard format nor a set of lightweight tools that allow small laboratories or individual investigators to share imaging and *meta-data* in a structured way, nor a set of tools that allows for queries across existing databases or data sharing efforts (Poline et al., 2012).   These problems are especially acute when attempting to construct large datasets from data available through online repositories, each of which uses different structures of storing meta-data.

The options for making data available are limited to putting raw data sets online, or putting raw and processed data sets online together with descriptions of the derived data in text documents (or to be gleaned from accompanying published manuscripts, e.g.,. www.openfmri.org). While these simple approaches make data available to the community, they generally provide limited meta-data thought to be of interest to the community by those supplying the data, they provide meta-data in unstructured formats making it difficult to use computationally, and they do not provide a framework to gain access to the data where it is hosted which often has much richer meta-data and/or may contain important ancillary information. Sharing data with an agreed upon structure directly from queryable data resources would not only provide a richer set of meta-data to investigators, it would lessen the burden in obtaining data from multiple resources while also making the data available programmatically.

In the domain of brain imaging, there is an implicit difference between data and meta-data. The primary distinction comes from considering binary forms of information as data and information related to the provenance of, generation of and associated with such binary forms as "meta-data". However, from an information perspective there is no difference between these two terms.  Current brain imaging databases and software attempt to capture "meta-data" or more textual information in one of three ways: 1) very explicit and relational data structures (e.g., XCEDE, XNAT schemas); 2) headers of binary file formats (e.g., DICOM, MINC2, NIFTI header extensions) or 3) human readable log files (e.g., FSL's FEAT, FreeSurfer recon-all). In each of these cases, there are no general mechanisms for querying such information.  For example, in MINC 2.0, a global mandatory attribute called 'history' (a character array) is used to implement an audit trail. According to MINC guidelines (Vincent et al. 2004), "All MINC applications should append a line containing: date, time of day, user name, program name and command arguments for each processing step." Similarly in AFNI (afni.com), FreeSurfer (surfer.nmr.mgh.harvard.edu) and FSL (www.fmrib.ox.ac.uk/fsl), such information are encoded in log files. Without a formal structure or binary readers, such information becomes only useful for manual visual inspection. Furthermore, these binary file formats or log files require special software to first extract the information and store it in a database before queries can be done. In our model, presented in this manuscript, we depart from these approaches by using a relatively simple data model (very few conceptual elements) that can encode a variety of data, including but not limited to phenotypic information, provenance of processes and the users and organizations involved. The data model provides the basic elements that can be used to build object models or scaffolds capable of storing many different forms of brain imaging data. Furthermore, the information can be exported using standards such as RDF and enabling other standards such as SPARQL (www.w3.org/TR/rdf-sparql-query) to perform queries that would in general be hard to do in the context of relational databases.

1 Derived Data Working Group

Gaining support for a new standard within the community is difficult.  In evaluating data format standards for storing medical images (e.g. DICOM, NIfTI) we find that often the most pervasive and well-supported formats are those that have been engineered through a community effort. The Neuroimaging and Technology Initiative's data format working group (NIfTI; nifti.nimh.nih.gov) is a notable example.  The organizers, under the direction of the National Institute of Mental Health (NIMH), brought together both domain-specific software developers and those involved in large neuroimaging consortia to address problems with sharing clinical images from a data format perspective.  The format was successfully adopted by most neuroimaging software providers and helped to standardize the way in which researchers accessed binary image data.  NIfTI did not however provide a format for storing rich image meta-data and data provenance information, both of which are critical for using shared data.

In response to the need for structured documentation of derived data in the context of neuroimaging, a derived data working group was formed (https://wiki.birncommunity.org/display/FBIRN/Derived+Data+Working+Group).  The group began in 2010 as a grassroots effort between members of the Function Biomedical Informatics Research Network (FBIRN) and focused on sharing analysis results within that consortium by expanding on previously developed FBIRN tools for the sharing of raw neuroimaging data.  It was quickly understood that the need for a structured way of documenting derived data in neuroimaging was a problem common among many scientists in the field and the effort was subsequently supported by the Biomedical Informatics Research Network Coordinating Center (BIRN-CC) through the formation of a sanctioned working group.  The working group expanded and gained traction in 2011 by the participation of members in the International Neuroinformatics Coordinating Facility's (INCF) Neuroimaging Standards for Datasharing Taskforce (www.incf.org/core/programs/datasharing).  It became obvious that many in the international community also shared the need for structured documentation of derived data and provenance in the service of sharing neuroimaging datasets.  The inclusion of the INCF participants resulted in an expanded focus, beyond strictly augmenting FBIRN tools, with the result of generalizing the core ideas of those tools to address a broader audience.

Today, the working group participants receive financial support from both the INCF and the BIRN as well as other funding sources; yet their contributions to the derived data work are predominantly voluntary.  The working group has weekly calls, open to anyone interested in contributing, and continually interacts with the INCF and BIRN-CC. Further, the INCF data sharing task force has multiple face-to-face meetings per year where the contributions from the derived data working group are shared and feedback elicited from the larger international community. The current focus of the working group is on the components for structured data sharing discussed in section 3 and the inclusion of appropriate sub-components into both common analysis tools and existing database resources.

In this manuscript, we present our work on facilitating the structured interchange of neuroimaging meta-data and our progress towards a unified set of definitions and tools for data and meta-data exchange. We focus not on the binary image data, but on the image meta-data and associated provenance in the service of facilitating distributed access to shared neuroimaging data. The work is applicable to both tool developers and end users and addresses two main questions a researcher has when contemplating the public sharing of their data. 1) How much, and what specific, meta-information is required about my data to allow effective sharing of it? 2) What format should I use to document my meta-data? The remainder of this manuscript is organized as follows. In section 2 we discuss why sharing neuroimaging derived data is important and why a structured approach to data provenance is necessary. In section 3 we present the components of our framework to facilitate enhanced data sharing through structured meta-data and provenance. In section 4 we discuss how the components from section 3 can be used by database mediation technologies to improve efficiency. In section 5 we discuss how the components from section 3 can be used in knowledge engineering technologies to represent and share statistical findings from experiments in a standardized way. Finally, in section 6 we discuss the current state of the work and future directions.

2. Importance of Derived Data Documentation

With any dataset, and in particular with neuroimaging data, there are many required initial steps both to organize and to prepare the data for analysis. With neuroimaging data, there are a series of image processing steps required before any statistics are performed. These steps can include image registrations, transformations and filtering operations and are non-trivial in their selection and implementation (Keator et al., 2009). These pre-processing steps are common for all neuroimaging data modalities, be it Positron Emission Tomography (PET) or structural and functional Magnetic Resonance Imaging (MRI). After pre-processing, in dynamic imaging modalities, voxelwise time-series analysis methodologies are used to infer task-related changes in the brain on an individual basis. The choice of statistical models and knowledge of parameters used is imperative for both the interpretation of the individual subject results and for evaluating the appropriateness of including subjects in higher level group analyses.

For researchers without significant experience in medical imaging analysis or access to significant computing resources, working with raw data, can present several challenges. Initially, images must be acquired from colleagues and/or various repositories. The data is often poorly documented and may be stored in different data formats. After download, the end-user needs to perform pre-processing steps which involves significant computational resources and expertise in using specialized software packages (within which the proper choice of input parameters is often unclear). Given these issues, many users prefer to work with derived data, i.e., data to which pre-processing algorithms have already been applied. In order for the sharing of derived data to be feasible the steps taken to arrive at the derived data, its 'provenance', must be adequately documented.

Data provenance refers to the history of every interaction with a piece of data, beginning at the time of its acquisition. In the case of neuroimaging data, this includes descriptions of the imaging hardware and parameters used in the acquisition data (Mackenzie-Graham et al., 2008). Many imaging hardware vendors, though not all, output data in the Digital Imaging and Communications in Medicine (DICOM) format, the standard for handling storing and transmitting medical imaging data. While DICOM captures much of the necessary meta-data, the location of this information can vary across manufactures and is often hidden in private sections in the DICOM header. Although DICOM is fairly complete, it is not well supported in the research community during the data analysis phase. Standard practice is to convert the original DICOM files into more concise image formats with primarily spatial meta-data (e.g. NIfTI). Such lossy transformations generally result in missing meta-data in the derived products. The retention of meta-data and the capture of pre-processing and analysis provenance has become an important topic in the neuroimaging field and is currently without any standards. The XCEDE XML schema has been useful in providing a structure for capturing meta-data associated with neuroimaging experiments but has fallen short in providing complex and easily extensible structures for derived data, meta-data, and provenance. One of the aims of this work is to provide tools and techniques for retaining as much meta-data and provenance as possible in an extensible framework such that these data can accompany derived data sets.

3. Components for Structured Derived Data Sharing

The following sections describe the components of our framework for structured derived data sharing and documentation. The components include (1) a *structured terminology* that provides semantic context to data, (2) a *formal data model* for neuroimaging with robust *tracking of data provenance*, (3) a web service-based *application programming interface* (API) that provides a consistent mechanism to access and query the data model, and (4) a provenance library with neuroimaging extensions that can be used for the extraction of provenance data by image analysts and imaging software developers.

3.1 Terminology

For the sharing of data to be useful, the data must not only be stored in an organized fashion, but meta-data that captures contextual information about how the data was acquired, processed, and analyzed, must also be made available to the prospective user. In addition, the meta-data must describe the data using terms that are unambiguously defined. Unambiguous definitions of terms are necessary for researchers in order to produce meaningful results when combining data from disparate sources. For example, if you don't know what "TE" means and/or you are not sure that another data set means the same thing when it uses "TE", then it's not clear that the data can be combined. Providing an explicit and web accessible definition with data type information (e.g., string, integer, etc.) allows data providers to be clear about their intentions. There are a number of lexicon development efforts underway that are germane to brain imaging research and the neuroscience community. Two lexicons that fit well with our efforts are NeuroLex (Larson et al., 2011) and RadLex (Langlotz, 2006), which are curated terminology resources for the domains of neuroscience and radiology, respectively.

One commonly encountered scenario is the existence of data and tool collections that have a defined schema and/or a fixed set of describing terms, but do not provide definitions for these terms. Therefore, though these collections make data or tools available along with associated meta-data, users may not know precisely what is meant by each meta-data term.  For example, a data set may include "nonlinear registration" in its provenance, but with no additional qualifying information the user will not be able to determine further attributes of the registration algorithm that could profoundly affect the data.  In general, the lack of standardized terms makes it difficult for query tools to search across collections at different institutions and for users to precisely know the attributes of the data they discover.

The goal of our terminology work is to provide definitions for terms used at each stage of the data lifecycle of imaging experiments and to curate the terms within two lexicons (NeuroLex and RadLex).  The source of the terms are 1) the XML-Based Clinical Experiment Data Exchange schema (XCEDE) XML schema (Gadde et al., 2012), 2) public DICOM fields (http://ushik.ahrq.gov/, http://medical.nema.org/), 3) private, vendor-specific DICOM fields in cases in which they are known by the community contributing to the terminology, 4) query-related terms from the NITRC database (Buccigrossi et al., 2008), and  5) terms used in BIRN's Human Imaging Database (HID) (Ozyurt et al., 2010).  Each term is provided with a definition and then placed within NeuroLex or Radlex by comparing the term with the lexicon's existing structure and adding parent terms and definitions where necessary.

We have focused our initial efforts on terms used in MR-based imaging protocols, due to the existence and wide adoption of the DICOM standard and the increasing availability of data and tools that use this standard.  It is particularly important to define and add terms that describe data stored in the private fields allowed by the DICOM standard because, although the private fields are heavily used by vendors, there is no consistent nomenclature among vendors in use for the private fields.  We have chosen terms in use by XCEDE, DICOM, NITRC, and HID because these sources include the terms that have current or increasing use in the field.  This work moves us closer to the goal of having a set of well-defined standardized terms that can be used in the sharing of imaging data.  Although the terminology was created in conjunction with the other components presented in sections 3.2 - 3.4, it stands on its own and can be used independently.

3.2. Neuroimaging Data Model (NI-DM)

The NI-DM is an ongoing effort to develop an extensible model for neuroimaging derived data with support for rich, queryable provenance and metadata.  The initial developments were based on extending the XCEDE XML schema (Gadde et al., 2012) in a technology agnostic manner. The XCEDE XML schema is a data exchange format for encoding general research data and meta-data. The schema allows for storing information in the context of a flexible and extensible experiment hierarchy, accommodating arbitrary configurations centered around *Project*, *Subject*, *Visit*, *Study*, *Episode*, and *Acquisition* objects, as well as limited information about data provenance. By representing many of the common types of information found in neuroimaging databases, XCEDE can facilitate data integration and act as a common data model, or mediated schema (Louie et al., 2007), that captures information from heterogeneous systems in a common XML syntax. In this way, available database resources can be described using a common language that simplifies data integration and sharing efforts, much in the same way NIfTI simplified imaging data exchange across analysis platforms.  While the XCEDE schema was appropriate for documenting neuroimaging datasets collected during the course of an experiment, it is unsuited for modeling and querying across complex derived data created from many of today's workflow systems.  Often derived datasets are created from multiple processing streams forming directed graphs.  XCEDE was not designed to support such complex workflows, supporting limited provenance encoding.  As implemented, even using XPath technologies, it would be difficult to perform queries like "Return the cortical volumes of participants under the age 15 processed with FreeSurfer version 5 or higher." By explicitly linking such information together in a formal manner, such queries allow one to navigate the data space, retrieving and filtering relevant information more efficiently.

Our approach with NI-DM is to capture, in a single data model, the different components of a research activity (e.g., participants, researchers, software, and hardware) and their relations. With the observation that activities represent transformations of data, it is easy to conclude that data and provenance are intricately linked and should be captured together. Figure 1 provides a high level view of how NI-DM can be used to explicitly model research activities in the context of provenance, which provides a computable representation that supports rich query capabilities.

As an example, consider the following query: "Find all participants under 18 years of age with a left putamen volume greater than 6,000 mm$^3$ that was calculated using FreeSurfer." This query requires access to research activities located in the *project information, workflow information, and derived data information* sections of Figure 1. The *project information* section includes participant demographics, while *workflow information* describes the software and computational activities, and *derived data information* lists all the corresponding output statistics. In all these cases, it is imperative that we, as a community, identify common data and information elements that can be used to annotate our resources in an unambiguous and consistent manner, highlighting the need for published terminologies (section 2). Additional example queries that information encoded with NI-DM can address: (1) Find the gray matter volume of all limbic structures for participants with Autism between the ages of 8 and 14; (2) Find the cortical volumes of participants under the age 15 processed with FreeSurfer version 5 or higher; (3) Find the cluster coordinates of peak T-statistic face-object contrast which were realigned with SPM and smoothed with a kernel greater than 6mm; (4) Find the diagnosis and caudate volumes of all subjects with a MMSE score less than 26 that were processed with FSL FAST version 4 or higher.

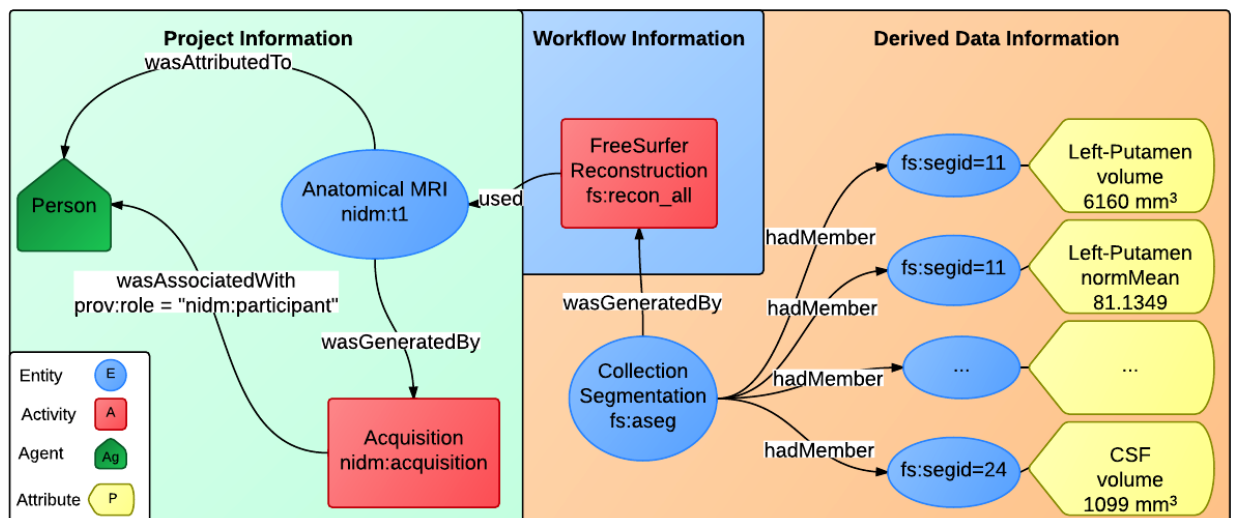

Figure 1: NI-DM can bridge information dimensions across Project, Workflow, and Derived Data. The "nidm" and "fs" namespaces are used to reference terms or annotations specific to NI-DM or the FreeSurfer analysis package, respectively. Explicit relationships link components together blurring the line between project information and processing workflows.

One of the primary design goals of XCEDE was to allow for *a priori* and *post hoc* documentation of acquired data and data analysis provenance such that future users of the data would have enough information to understand, at a high level, the data contained within the XML file. In NI-DM these goals are extended to allow for derived data *replication* using precisely the same model components as used to document the acquired data. The NI-DM components that enable such documentation are:

1. Provenance - capturing a description of a process, such as when data are derived from other data (e.g. as a result of analysis). Provenance elements are used to store the step-by-step execution details of a process or processes that generated derived data.

2. Terminologies - all data elements can be annotated with terms from ontologies and/or lexicons, such as NeuroLex or RadLex. These links to external information resources provide the meaning of a given value and act as a mechanism to later correlate, integrate, and reason over data using semantic web technologies.

3. Other annotations - all core NI-DM components can be annotated, either with free-form text labeled by an author, pointers to external resources (such as publications), or user-extensible typed annotations with an appropriate terminology and/or ontology reference, specifying the meaning of the annotation.

3.2.1 Provenance and the W3C PROV Standard

Data provenance is a description of how an artifact (including digital information) comes into existence. This description is dictated by the agent(s), entity(s), and activity(s) that were used to construct some piece of data or information. The concept of provenance is generic in nature and thus applicable across a broad range of domains, including neuroimaging.

The Provenance Working Group (w3.org/2011/prov) of the World Wide Web Consortium (W3C) designed PROV, a suite of specifications, to address the need for a generic approach to representing provenance information. PROV evolved from several years of theoretical and applied provenance research initiated by a series of Provenance Challenges that took place from 2006 to 2010, with fMRI provenance as a driving use-case (Moreau et al., 2008) (twiki.ipaw.info/bin/view/Challenge/). Out of the Provenance Challenge came the Open Provenance Model (OPM) (openprovenance.org) which was implemented and evaluated in a number of production workflow systems (Hull et al., 2006). The lessons learned from implementing OPM were used to drive the design of PROV and work toward a standard W3C recommendation for the representation of provenance. Our decision to adopt PROV is an effort to foster interoperability within and beyond the domain of neuroimaging, which is justified by the years of applied research in the Provenance Working Group.

Although a complete overview of PROV is outside the scope of this discussion, here we introduce key concepts needed to understand NI-DM. The suite of PROV specifications provides an explicit representation of provenance semantics in the PROV Data Model (PROV-DM) (http://www.w3.org/TR/prov-dm/) document. PROV-DM describes the core elements and relationships that can be used to represent any process as a directed graph by linking three structures (entities, activities, and agents) using a specific set of relationships as shown in Figure 2. PROV-DM is defined in a technology agnostic way, such that it is not tied to the limitations of any particular language. This allows for developers implementing PROV to select from any number of data formats, including XML, JSON, and RDF. With such a flexible and generic approach to representing provenance, additional constraints are needed to make the PROV specific for a given domain. PROV-DM provides extension points for including domain specific semantics, which we discuss in the following section.

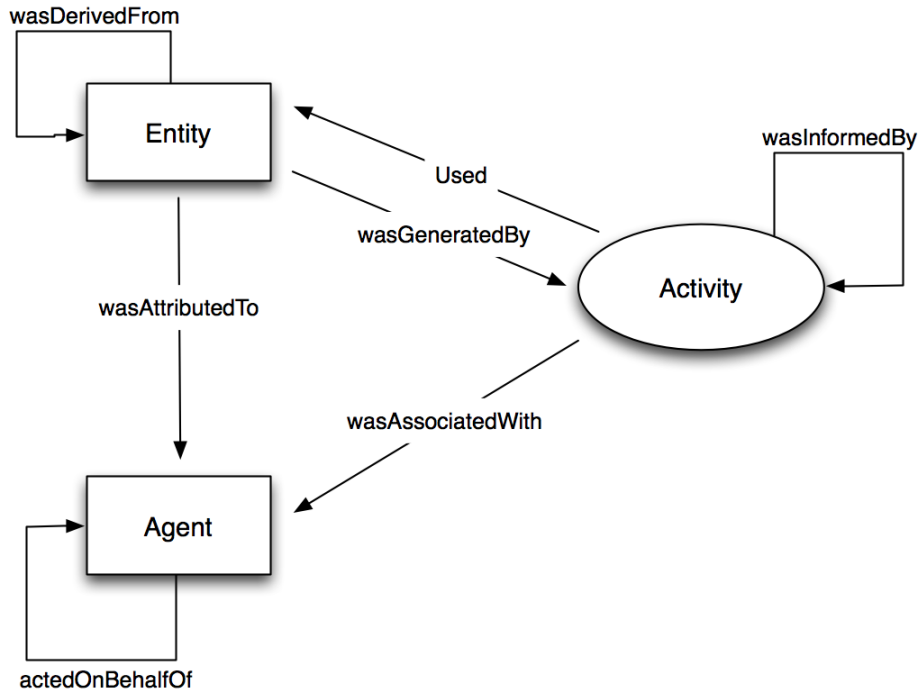

Figure 2: PROV-DM Core Structures are 1) *entity* - a physical, digital, conceptual, or other kind of thing with some fixed aspects, and can be real or imaginary., 2) *activity* - something that occurs over a period of time and acts upon or with entities; it may include consuming, processing, transforming, modifying, relocating, using, or generating entities., 3) *agent* - something that bears some form of responsibility for an activity taking place, for the existence of an entity, or for another agent's activity, and the relationships a) wasDerivedBy, b) used, c) wasGeneratedBy, d) wasInformedBy, e) wasAssociatedWith, f) actedOnBehalfOf, g) wasAttributedTo. (figure adopted from http://www.w3.org/TR/prov-dm/ ).

PROV itself is totally agnostic about neuroimaging, but provides simple structures for describing where and how new things come into existence. All of the simple data structures that PROV provides are exemplified in the fMRI use case, which relies heavily on computational workflows that generate derived data using a set of inputs that can then be used as input to further processing.

3.2.2 Extending PROV-DM with XCEDE Constructs

PROV-DM's format-independence, natural representation of workflow, and extensibility points motivated the Derived Data Working Group to harmonize the model with the XCEDE XML schema. Our harmonization effort revealed that a PROV-DM compliant description of data provenance was highly redundant with some information already modeled in XCEDE XML. This observation led us to explore the use of PROV-DM extensibility points to model the XCEDE experiment hierarchy. By modeling the conceptual information contained within the XCEDE experiment hierarchy and relaxing the constraint on the strict hierarchical relationship embedded within the XCEDE schema, we found that both the raw and derived data, along with provenance could be modeled with the same basic set of structures and relationships. By adding specific extensions to make the core PROV-DM model more amiable to neuroimagers, the resulting model, NI-DM, was conceptually simpler to understand than XCEDE (three core structures and some relationships; Figure 2) and would solve our difficulties with XCEDE as discussed in section 3.2.

3.2.2.1 Data Integration and Harmonization Example

The section below shows an example in which NI-DM captures clinical and imaging meta-data (i.e., data and associated provenance information) from two neuroimaging databases. In our example we use the Human Imaging Database (Ozyurt et al., 2010) and eXtensible Neuroimaging Archive Toolkit (Marcus et al., 2007) to highlight the use of provenance, terminologies, and other annotations. The HID and XNAT management systems used in this example have been successfully integrated in the past using database mediation technologies (see section 4) therefore making them a particularly relevant example. The data used in this example reflects real data from the FBIRN data mediation project (Ashish et al., 2010). The NI-DM graph consistent with this example is shown in Figure 3, while each step in the example is written using the PROV Notation syntax developed by the W3C Provenance Working Group (a description of the PROV Notation can be found at http://www.w3.org/TR/prov-n/). The data described concerns two perspectives (i.e., HID and XNAT) on a participant's handedness data obtained by a neuropsychological test administrator and an anatomical MRI obtained by a radiology technician. The forms and scanners used to collect this data are heterogeneously represented in HID and XNAT. The data are encoded with a varying level of semantics that ranges from free-text to explicitly defined using Unique Resource Identifiers (URIs). Data consumers (human or machine) can visit the URIs to obtain further information about the instruments used and determine the exact nature of how the data was obtained. Similar descriptions can also accompany processed imaging data (for an example see section 3.4).

In this example, we start by looking at the representation of two clinical questionnaires, which can be thought of as a protocol or plan for acquiring data and not the filled out form itself. We define two entities, one for each questionnaire from the respective databases:

```
entity(plan_1,[prov:type='prov:Plan',
            prov:type='neurolex:Handedness_Form',
            prov:type='hid:Edinburgh_Handedness',
            prov:label="Subject Handedness Form",
            nidm:url="http://myform.com/Edinburgh.pdf"])

entity(plan_2,[prov:type='prov:Plan',
            prov:type='neurolex:Handedness_Form',
            prov:type='xnat:Handedness',
            prov:label="Subject Handedness Form",
       nidm:url="http://myform.com/Handedness.html"])
```

These entity definitions contain an identifier ("plan_xx"), followed by five attribute-value pairs used to qualify the entities: (1) prov:type - which indicates a term defining the entity type(s), with the first value being of type "prov:Plan" that indicates these entities represent a set of actions or steps intended by one or more agents to achieve some goals, (2) prov:type - with a value referencing an external term in NeuroLex, (3) prov:type - a reference to the field name at the data source, (4) prov:label - which provides a human readable label for user interfaces, and (5) "nidm:url" - which indicates the URL (Uniform, or universal, resource locator) where this entity is accessible on the Web. This example demonstrates the flexibility in extending one of PROV's core structures by referencing different namespaces (e.g., hid, neurolex, nidm, prov, xnat etc.) where additional information can be gleaned. The multiple "prov:type" tags enables elements to be semantically annotated, thus providing a mechanism to map a given field to common data elements. Next, we define the "activity" of acquiring the questionnaires, as well as anatomical MRI scans:

***Activity for determining handedness in HID:***
activity(acquisition_1,
      2001-01-01T00:00:00,
      2001-01-01T00:15:00,
       [prov:type='nidm:acquisition',
        prov:type='neurolex:Handedness',
        prov:type='hid:Edinburgh_Handedness'])

***Activity for determining handedness in XNAT:***
activity(acquisition_2,
      2001-01-01T00:20:00,
      2001-01-01T00:30:00,
       [prov:type='nidm:acquisition',
        prov:type='neurolex:Handedness',
        prov:type='xnat:Handedness'])

***Activity for acquiring a T1 weighted MRI in HID:***
activity(acquisition_3,
      2001-01-01T00:00:00,
      2001-01-01T00:15:00,
       [prov:type='nidm:acquisition',
        prov:type='neurolex:T1',
        prov:type='hid:spgr'])

***Activity for acquiring a T1 weighted MRI in XNAT:***
activity(acquisition_4,
      2001-01-01T00:20:00,
      2001-01-01T00:30:00,
       [prov:type='nidm:acquisition',
        prov:type='neurolex:T1',
        prov:type='xnat:mprage'])

These activity definitions contain an identifier ("acquisition_xx"), start timestamp, end timestamp, and three "prov:type" attributes. In the above examples, the first "prov:type" attribute references a NI-DM type derived from the XCEDE experiment hierarchy, while the second defines the specific type of acquisition (i.e., handedness questionnaire or T1 weighted MRI), and the third highlights the HID and XNAT namespaces for the given activity. Additional annotations can also be defined. Next, we can define agents:

***Agents of type Person from HID:***
agent(person_1,
         [prov:type='prov:Person',
          prov:label="Person"])

agent(person_2,
         [prov:type='prov:Person',
          prov:label="Person"])

***Agents of type Person from XNAT:***
agent(person_3,
         [prov:type='prov:Person',
          prov:label="Person"])

agent(person_4,
         [prov:type='prov:Person',
          prov:label="Person"])

These agent definitions contain an identifier ("person_xx"), and two attribute-value pairs used to qualify the agents: (1) prov:type - which has the value of "prov:Person", and (2) prov:label - which provides a human readable label. These attribute-value pairs list a limited amount information about an agent.  Alternatively, a richer set of attributes could be included here providing additional context about the agents. Next we look at each of the "wasAssociatedWith" relations that link activities, agents, and plans:

***Associate HID plan with handedness acquisition and plan:***
wasAssociatedWith(wAW_1, acquisition_1, person_1, plan_1,
                  [prov:role='neurolex:NP_Test_Administrator'])

wasAssociatedWith(wAW_2, acquisition_1, person_2, plan_1,
                  [prov:role='neurolex:Participant'])

***Associate XNAT plan with handedness acquisition and plan:***
wasAssociatedWith(wAW_3, acquisition_2, person_3, plan_2,
                  [prov:role='neurolex:NP_Test_Administrator'])

wasAssociatedWith(wAW_4, acquisition_2, person_4, plan_2,
                  [prov:role='neurolex:Participant'])

***Associate HID anatomical MRI acquisition:***
wasAssociatedWith(wAW_1, acquisition_3, person_1, -,
                  [prov:role='neurolex:Radiology_Technician])

wasAssociatedWith(wAW_2, acquisition_3, person_2, -,
                  [prov:role='neurolex:Participant'])

***Associate XNAT handedness MRI acquisition:***
wasAssociatedWith(wAW_3, acquisition_4, person_3, -,
                  [prov:role='neurolex:Radiology_Technician])

```
wasAssociatedWith(wAW_4, acquisition_4, person_4, -,
                  [prov:role='neurolex:Participant'])
```

The above definitions for the "wasAssociatedWith" relationship contain an identifier ("wAW_xx") that is used to relate an activity ("acquisition_xx") to an agent ("person_xx") based on a given (optional) plan ("plan_xx"), and defines a single attribute-value pair to identify the role of agents in the activity: (1) prov:role - which has a value from NeuroLex further qualifying the role played by an agent for that specific activity. Each set of wasAssociatedWith relations indicate different roles for each agent, which enables the ability to (optionally) denote responsibility for some activity occurring (i.e., not just the participant data but whom collected the data). Next we will look at the entities that these activities, agents, and plans produce.

***Entities from handedness activities:***
```
entity(value_1,[prov:type='neurolex:Handedness',
                prov:type='hid:Edinburgh_Handedness',
                prov:label='Handedness',
                prov:value='neurolex:right_handed'])

entity(value_2,[prov:type='neurolex:Handedness',
                prov:type='xnat:Handedness',
                prov:label='Handedness',
                prov:value='neurolex:right_handed'])
```

***Entities from T1 acquisition activities:***
```
entity(value_3,[prov:type='neurolex:T1,
                prov:type='hid:spgr',
                prov:label='T1',
                prov:value='http://fbirnbdr.nbirn.net/T1.nii.gz'])

entity(value_4,[prov:type='neurolex:Repetition_Time,
                prov:type='hid:tr',
                prov:label='Repetition Time',
                prov:value='2.0'])

entity(value_5,[prov:type='neurolex:T1,
                prov:type='xnat:mprage,
                prov:label='T1',
                prov:value='http://central.xnat.org/T1.nii.gz'])

entity(value_6,[prov:type='neurolex:Repetition_Time,
                prov:type='xnat:tr',
                prov:label='Repetition Time',
                prov:value='2.0'])
```

These entity definitions represent the result of the above activities of acquiring data from a Handedness questionnaire. Each entity has an identifier (value_xx or location_xx) and three attribute-value pairs used to qualify the entities: (1) prov:type - which identifies a term in NeuroLex that database specific terms can be mapped to, (2) prov:type - which identifies a source specific term (i.e., HID or XNAT), (3) prov:label - a human readable label, and (4) prov:value - which contains the actual value acquired. The handedness section only includes a single value from each neuroimaging database, while the T1 acquisition has two values for each database. Next we define a collection that is used to group together related entities, like those from the "T1 Acquisition":

 entity(collection_1,[prov:type='prov:Collection',
                                    prov:type='neurolex:T1,
                                    prov:type='hid:spgr',
                                    prov:label="T1 Parameter Collection"])

 entity(collection_2,[prov:type='prov:Collection',
                                    prov:type='neurolex:T1,
                                    prov:type='xnat:mprage',
                                    prov:label="T1 Parameter Collection"])

This entity definition represents a collection of entities from a "T1 Acquisition", with an identifier (i.e., collection_1, collection_2) and four attribute-value pairs, (1) prov:type='prov:Collection' - which identifies this entity as a Collection type, (2) prov:type='neurolex:T1' - which identifies this entity as a collection of entities related to the "neurolex:T1", (3) prov:type='hid:spgr' or prov:type='xnat:mprage' - which defines , (4) prov:label - which provides a human readable label. Next we can group together "neurolex:T1" entities using the hadMember relationship:

***Collection of HID T1 Acquisition Meta-data:***
hadMember(collection_1, value_3)
hadMember(collection_1, value_4)

***Collection of XNAT T1 Acquisition Meta-data:***
hadMember(collection_2, value_5)
hadMember(collection_2, value_6)

The hadMember definitions connect the collections to each of their member entities using the entity identifiers (i.e., value_xx). We close this set of examples with the wasGeneratedBy relationship:

***Handedness activities and resulting entities:***
 wasGeneratedBy(value_1, acquisition_1, 2001-01-01T00:30:00)
 wasGeneratedBy(value_2, acquisition_2, 2001-01-01T00:30:00)

***T1 Acquisition activities and resulting entities:***
 wasGeneratedBy(collection_1, acquisition_3, 2001-01-01T00:15:00)
 wasGeneratedBy(collection_2, acquisition_4, 2001-01-01T00:15:00)

The wasGeneratedBy definitions above link both collections of values (i.e., collection_1 and collection_2) or specific values (i.e., value_1, value_2) to the relevant activities and the time that they were completed. The above example demonstrates how NI-DM represents core structures from the XCEDE experimental hierarchy (Subject, Acquisition) within the context of provenance (figure 2).  Further, it demonstrates the flexibility of documenting provenance with NI-DM while still providing semantically meaningful annotations.

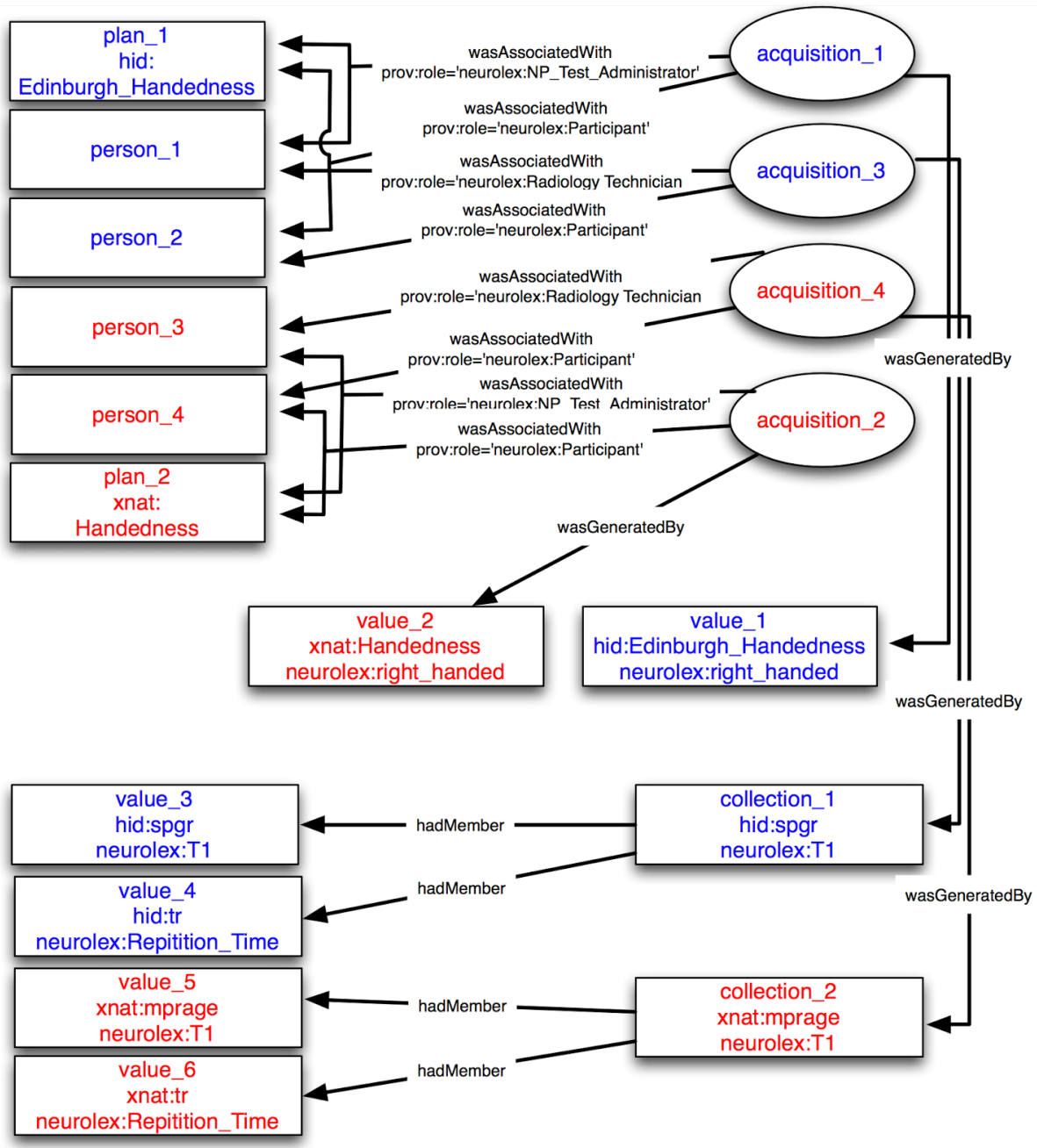

Figure 3: Provenance graph of the NI-DM example in section 3.2.2.1. Entities are represented by rectangles and activities as ellipses. Associations are indicated with edge labels. Text has been color coded to indicate which data source (hid, xnat) the item is associated with.

3.3 Web Services

While individual neuroscience databases provide mechanisms to query and download information within a given framework (e.g., Allen Institute, COINS, HID, IDA, LORIS, NIMS, XNAT), there is no standardized way to programmatically access information stored in these heterogeneous systems. Developing a lexicon (3.1) and data model (3.2) focused on neuroimaging provides the standards needed to create a common data exchange layer. By mapping a core set of information from individual neuroimaging databases into this framework, database providers can publish information about their resources in a common way. A web service application programming interface (API) to access and query shared neuroimaging data will enable the development of interoperable client applications capable of consuming resources available across disparate brain imaging data management systems.

To meet this end, we are developing an API for providing uniform access to neuroimaging databases. Conceptually, the API is a service for accessing the NI-DM core structures: (1) activity (e.g., acquisition), (2) agent (e.g., subject), (3) entity (e.g. resources), and their relationships (e.g., subject *wasAssociatedWith* acquisition). Neuroimaging databases conforming to the API implement a mapping of their local resources to NI-DM and provide a mechanism to request resources. The API is not tied to a specific language or technology, but for web-accessible databases, the Representational State Transfer (REST) architecture is a natural fit. For example, a REST implementation of the API responds to a request for list of subjects by returning an NI-DM compliant representation in an available format (e.g., XML, JSON, and RDF).

The goal is to increase the ease of finding data relevant to a given project. This necessitates querying data in a scientifically relevant way. For example, currently it is not possible to query the BIRN HID, XNAT Central, and ADNI-LONI databases for all T1-weighted data sets, from healthy males between the ages of 18-38, processed with Freesurfer version 5.0.0. To facilitate such a query each data repository would either implement the NI-DM API or a mapping from an existing, custom API to the NI-DM API. Queries could then be issued programmatically (assuming appropriate privileges to query) or through a mediator interface to each of the data repositories. The resulting data would be returned in an available format (e.g., XML, JSON, RDF, etc.) consistent with the NI-DM specifications. Because the data model is un-ambiguous and has support for rich meta-data and provenance, the results of the query can be delivered to the user in a consistent, structured, and well documented way.

We acknowledge that the implementation of the API requires resources from the data providers. Currently each data provider is either without a web service layer or implementing their own APIs because the domain is without a standard. The benefits of a common API are undeniable but require a published, well documented standard that has sufficient flexibility and is supported by both data providers and tool developers. Given the structure of the Derived Data Working Group and the broad exposure through the INCF and BIRN groups, we have been able to gain traction in promoting this goal (see section 3.4). Discussions with the developers of many neuroimaging database and tool developers are helping to refine this specification which will speed up the first implementation of the API. Our goal is that in the near future, existing databases and those under development will implement this protocol and expose existing and newly acquired datasets in a common data access framework.

3.4 Provenance Library and Integration with Neuroimaging Tools

To facilitate broad use of the standards described in sections 3.1-3.3, both libraries for programmers and integration with popular neuroimaging analysis tools are necessary. The INCF's working group on data sharing and the Derived Data Working Group has jointly developed a C prototype library enabling easier construction of PROV-DM compatible XML formatted provenance representations for neuroimaging applications. The library has bindings to Python, JAVA, and Matlab and is freely available on the GitHub repository

(github.com/INCF/ProvenanceLibrary). Based on feedback from our community of collaborators and to maximize version synchronization with the core W3C PROV-DM, we have stopped developing the C library and focused our efforts on contributing to and using the JAVA ProvToolbox (github.com/lucmoreau/ProvToolbox) and the Prov Python library (pypi.python.org/pypi/prov), being developed by the W3C. Our contributions reflect neuroimaging domain specific extensions to the core PROV-DM model. The basic documentation for using these libraries are available here (openprovenance.org/java/site/prov/apidocs/) and we have examples of using these libraries for brain imaging here (nidm.nidash.org).

To incorporate support for provenance into popular neuroimaging analysis tools, the working group has begun developing scripts for SPM (www.fil.ion.ucl.ac.uk/spm/), FSL (www.fmrib.ox.ac.uk/fsl/), and FreeSurfer (surfer.nmr.mgh.harvard.edu/) image analysis software packages. For the SPM package, version 8, we have developed a tool that extracts processing steps and constructs an XML provenance file during the execution of the SPM batch processing software.

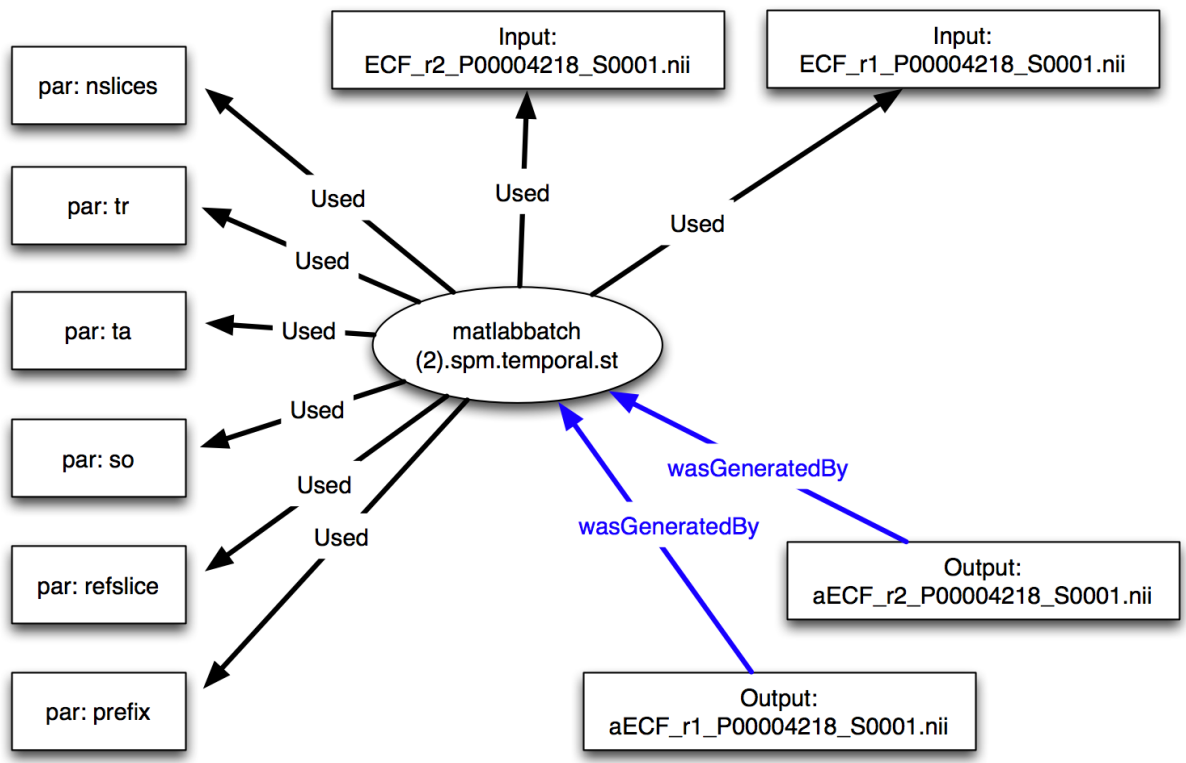

Figure 4: Provenance graph of a slice timing correction process on functional MRI data using the batch processing capabilities of SPM8. The provenance XML file is created using automated scripts run in Matlab under the SPM8 software. The graph is created to visualize the relationships between input entities and parameters entities (black edges) and the output entities (blue edges) with the activity entity represented by an ellipse.

The SPM8 batch processing software allows the user to create a series of analysis procedures once and use them repeatedly. Figure 3 demonstrates a graph created from the provenance XML generated by the extraction tool for SPM. Entities are represented by boxes and activities are represented by ellipses. The relationship *"Used"* and *"wasGeneratedBy"* between the entities and activities are encoded with edge-labels (arrows). In the provenance XML file, the activity (ellipse in Figure 3) is annotated with start and end times, a label, and a node id:

```
<prov:activity prov:id="a_1">
   <prov:startTime>07-Jun-2012 14:06:39</prov:startTime>
   <prov:endTime>07-Jun-2012 14:09:00</prov:endTime>
   <prov:label>matlabbatch{2}.spm.temporal.st</prov:label>
</prov:activity>
```

The entities are encoded in separate entries in the XML file, the entry for the parameter entity "tr" to the slice timing correction activity is given by:

```
<prov:entity prov:id="e_30">
   <prov:type xsi:type="xsd:string">parameter</prov:type>
   <ni:name xsi:type="xsd:string">par: tr</ni:name>
   <ni:value xsi:type="xsd:string">2</ni:value>
 </prov:entity>
```

Lastly, the relationship between the activity and the entity are encoded in an XML block whose provenance type is the relationship identifier:

```
<prov:used prov:id="u_20">
   <prov:activity prov:ref="a_1"/>
   <prov:entity prov:ref="e_30"/>
 </prov:used>
```

Similar outputs can be generated for many SPM processing streams, fully documenting the parameters to the process (activity) along with the inputs and outputs. With such rich provenance information, reconstructing and re-running the pipeline becomes a simple activity. We envision developing additional tools to create and execute an SPM batch process directly from the provenance XML files thus enabling researchers to both store complete provenance records in a compressed format and re-execute pipelines when needed.

For the FSL and Freesurfer packages the approach is very similar. Whenever an analysis is performed, the FSL/Freesurfer software automatically creates a log file. Our approach is to use this log file as input to our provenance extraction tool. This log file is parsed and the XML is created. The end result, regardless of whether SPM, Freesurfer or FSL was used, is an XML file storing the provenance information from either analysis in the same format.

To provide native support within the SPM, FSL, and FreeSurfer packages for provenance, the working group has reached out to the developers of each software package. The FSL group has tested the prototype provenance library to write XML meta-data directly to the NIfTI image headers in the extension area as images are moved through an FSL processing pipeline. The SPM and FreeSurfer developers have further volunteered to evaluate incorporating provenance within their future releases, acknowledging the need for such meta-data. Members from the Derived Data Working Group continue dialog with each development team to both encourage a unified provenance representation and to gather requirements and feedback on both the provenance library and the components discussed in sections 3.1-3.3.

4. Simplified Database Mediation

Information mediation (Adali et al., 1996; Arens et al., 1993; Draper et al., 2001; Florescu et al., 1996; Hammer et al., 1997; Lenzerini, 2002; Thakkar et al., 2007; Ullman, 2000) is an established data integration technology that can be used to provide integrated access to data in multiple, distributed, and possibly heterogeneous repositories. Developing a new information integration application using a mediator is a reasonably involved process but can be summarized in four main steps: 1) requirements gathering, 2) data modeling, 3) data source wrapping, and 4) final data source integration.

In requirements gathering the application developer(s) meet with domain experts to obtain an understanding of the need for data integration in their domain, the data sources they require integrated access over, and the kinds of capabilities they would expect from integrated data access. After requirements gathering, the application developer acquires an understanding of the particular data sources to be accessed. This involves understanding the type, interfaces (if any), content, and access information for each data source to be integrated. If the data sources are databases then the application developer must understand the database schema, how the data is represented in the schema, and how to query the underlying representation. The next step is data source wrapping. This involves configuring or developing "wrappers" where the wrapper is essentially a piece of software that provides a translation from/to the query language of the mediator and a data source that provides structured querying, or an abstraction over a data source that does not provide structured querying. For example, if the source is a relational database, a wrapper consists of software to translate the syntax of the mediator application to an SQL query on the underlying database schema. The last step is data source integration. In data source integration, the data source specific representations are mapped to an integrated data model using a set of integration rules created by the application developer in the course of understanding the individual data sources. Once these steps are complete, a mediator query interface can be created that provides the user with an intuitive, integrated way of querying and retrieving data from multiple resources.

The FBIRN data mediation project (Ashish et al., 2010) where we successfully achieved mediated integrated access to three disparate data sources, provides a direct context for considering the proposed NI-DM and API directions for neuro-informatics. In this project we integrated 3 data sources (i) The HID database, (ii) XNAT, and at a later stage (iii) SNPLims – a database of genetics data. XCEDE was used as the basis for the common "global" model in this application. Based on our experience we highlighted in Ashish et al., the steps of data source modeling, wrapping, and integration can be fairly time consuming for mediation application developers.

Going forward, the common model and API components presented in sections 3.1-3.3 can be leveraged to dramatically decrease the modeling, wrapping, and integration steps of mediation – both in terms of effort as well as the level of expertise required. If the data sources to be mediated implemented the web service API discussed in section 3.3, the mediator could query the data source using the functions and semantics defined in the API *without any additional work* on the part of mediation application developers. All data sources that implement the webservices API would be essentially queryable "for free" (given appropriate access permissions, security concerns, etc.), thereby eliminating the need for wrapper configuration or development. Further, because the API in section 3.3 is directly related to the NI-DM data model in section 3.2, the data source integration step could be reduced to just using the NI-DM data model itself. A mediation application developer, with the task to mediate between data sources that implement the API, would need to understand the specific requirements of the mediation activity, set up a mediator web interface that implements particular web service API function calls to the underlying data sources, and map those responses back to the internal integrated representation, the NI-DM. While effort is indeed required here, what is eliminated is the need to map data elements from each source into whatever global or unified model is used and in many cases define a unified model. Next, because the webservices API and NI-DM are integrated with the terminology in section 3.1, the mediator could provide *semantic annotations* consistent with the particular web service queries that are being supported by the information integration task. The combination of the common data model i.e., NI-DM and a congruent web service API for each source, essentially defines the data source model as well as the integrated "global" data model that would otherwise have to be developed.

Even with a common data model and standard APIs, effort is required to map each source to the common model. We can provide tools to alleviate some of the burden. The availability and employment of tools to assist data source providers in mapping their data to the common model has seen significant success in other clinical and medical data sharing initiatives. For example, the National Database for Autism Research (NDAR) has been successfully deployed using data from multiple organizations and repositories, mapping to a common data model. The Data Coordinating Center (DCC) in NDAR provides standard "templates", in the form of Excel spreadsheets, to data providers in which to provide their (autism) data. There are also format conversion tools provided to convert data elements such as data column names from the data source to the data elements in the common or "standard" data elements used by the DCC. Using such tools where appropriate and developing additional tools to aid researchers in marking up their data is an ongoing effort of the Derived Data Working Group.

5. Using Meta-data to Create Machine-readable Assertions

The development of standardized neuroimaging meta-data formats facilitates more than just data sharing: it could enable knowledge engineering technology to represent and share statistical findings from experiments in a standardized way. A usable meta-data standard can aid consumers of public data in properly analyzing and integrating the data with other datasets and studies. Currently however, when those analyses are published in a manuscript, all of the detailed provenance information is lost. The manuscript is, in a sense, a series of assertions; the experiment was performed, the data were analyzed, and the results support the concluding assertions. With proper meta-data being generated and included at each step along the way, the representation of these assertions could also be standardized into machine-readable and computable formats which could themselves become individual elements that form fragments of knowledge. One proposed format for these fragments is 'nanopublications' (Groth et al., 2010). This technology potentially allows for the combination of experimental assertions across many studies for automated reasoning, chaining hypotheses to results across neuroimaging studies, and potentially across scientific domains.

The standards being developed by the Derived Data Working Group described in sections 3.1-3.3 are being linked to other terminologies and ontologies where possible, to facilitate this larger vision. In particular (as described in a companion paper in this special issue), the data collection and analysis terms serve as variables in the Ontology of Experimental Variables and Values (OoEVV; www.isi.edu/projects/ooevv/). Thus they can be used to represent individual experimental findings generated by fMRI experiments as a form of nanopublication as provided by the Knowledge Engineering from Experimental Design (KEfED) framework, so that explicit assertions regarding cognitive processes, fMRI experiments, and the resulting brain activation patterns could be made available as computationally readable data (Russ et al. 2011). This is a first step toward formulating cognitive neuroscience in a computable system.

6. Conclusions

In this paper we have described our vision, the framework and its capabilities, along with initial progress towards the tools needed for structured sharing of raw and derived neuroimaging data across existing data resources. The sharing of meta- and provenance data is important when sharing raw imaging data, but becomes infinitely more complex and critically important when sharing derived data; even when such data is generated by highly standardized image processing pipelines (Gronenschild et al., 2012).

The practical challenges one faces when pooling data across neuroimaging resources include differences in data representations between databases, differences in terminologies, lack of provenance data due to lack of existing methods to generate provenance data and/or lack of uniformly stored provenance data, and the inability to query meta-data across imaging resources without significant and time-consuming data mediation efforts. Our Derived Data Working Group, supported by both the INCF and BIRN groups, is developing several practical tools to address these challenges along with libraries that take advantage of the primary deliverables. These practical tools include: a structured terminology that provides semantic context to data, a formal data model for neuroimaging with robust tracking of data provenance (NI-DM), a web service-based application programming interface (API) that provides a consistent mechanism to access and query the data model, a provenance library that can be used for the extraction of provenance data by image analysts and imaging software developers, as well as some scripts to extract provenance data from common analysis pipelines. In addition, we are reaching out to software developers with regard to having the libraries and unified provenance generation incorporated as part of existing software tools.

Similar to the exchange of binary imaging data between neuroimaging analysis software tools has benefited from the interaction of software tools with an agreed-upon unified format (e.g., NIfTI) along with the needed conversion tools, our vision is that the exchange of imaging meta-data between neuroimaging databases and other query mechanisms (e.g., pipeline tools), will benefit from an agreed-upon unified format for meta-data exchange along with an associated query interface and terminology.

The NI-DM and webservice API specification documents along with ongoing working examples produced by the Derived Data Working Group are freely available via the public Wiki: wiki.birncommunity.org/display/FBIRN/Derived+Data+Working+Group.  The work presented in this manuscript does not cover data access in terms of security. This is a topic that is addressed by the Security Working Group of the BIRN Coordinating Center ([www.birncommunity.org](www.birncommunity.org)) which provides services with regard to credential management, group policy management, and user registration and certificate authority. We continue to gather and include as much community input and support as possible as none of the meta-data sharing tools in development will be of use if they are not accepted by and do not serve the wider neuroimaging community.

In addition to accelerating scientific discovery by improving the ability to share neuroimaging data to test new hypotheses, the ability to share links to neuroimaging data, meta-data, and complete provenance data along with research publications is also likely to accelerate scientific discovery through earlier detection of non-replicable research findings; a topic that has recently received a lot of attention (Begley and Ellis, 2012 ; Yong, 2012). Such capabilities could accelerate scientific discovery by faster falsification of incorrect hypotheses based on such non-replicable research findings. Moreover, comprehensive data sharing will also provide a plethora of pilot data that other researchers can examine before deciding on their next research direction. This will likely result in more careful selection of research directions and a more efficient use of research funds. The ability to share data, meta-data, and provenance data will go a long way towards raising the standards for and improved rigor in neuroimaging research in that it provides a system for a "*stronger more transparent discovery process*" (Begley and Ellis, 2012). In general, but in particular with regard to acceleration of discoveries in the area of brain disorder research (e.g., 1mind4research.org), to which neuroimaging is most likely to make a large contribution, we want to spend the least amount of time moving in the wrong direction while chasing discoveries.

In conclusion, we believe that the framework and set of tools outlined in this manuscript have great potential for solving many of the practical issues the neuroimaging community faces when sharing raw and derived neuroimaging data across the various existing database systems for the purpose of accelerating scientific discovery.

Acknowledgements

This work was supported by the Function Biomedical Informatics Research Network (NIH 1 U24 U24 RR021992) and the BIRN Coordinating Center (https://www.birncommunity.org; NIH 1 U24 RR025736-01). The International Neuroinformatics Coordinating Facility provided financial support for some of the work reported in this article (to Satrajit Ghosh). This work was further supported by RC4 NS073008-01 (PI: Grabowski, co-author Nichols, B.N.) and the National Academies Keck Futures Initiative (PI: Grabowski, co-author Nichols, B.N).References

Adali, S., Candan, K.S., Papakonstantinou, Y., Subrahmanian, V.S., 1996. Query Caching and Optimization in Distributed Mediator Systems., SIGMOD Conference, pp. 137-148.